\documentclass[%
 reprint,
 amsmath,amssymb,
 aps,
]{revtex4-2}

\usepackage{graphicx}
\usepackage{dcolumn}
\usepackage{bm}
\usepackage{siunitx}

\begin{document}

\preprint{APS/123-QED}

\title{A Geometric Pathway for Tuning Ferroelectric Properties \\via Polar State Reconfiguration}

\author{Hao-Cheng Thong$^{1, 2}$}
\author{Bo Wu$^{3}$}
\email{Contact author: wubo7788@126.com} 
\author{Fan Hu$^{4}$}
\author{Pedro B. Groszewicz$^{4, 5}$}
\email{Contact author: pedro.groszewicz@helmholtz-berlin.de}
\author{Chen-Bo-Wen Li$^{1}$}
\author{Jun Chen$^{6}$}
\author{Mao-Hua Zhang$^{7}$}
\email{Contact author: zhangmh@wuzhenlab.com}
\author{Dragan Damjanovic$^{8}$}
\author{Ben Xu$^{9}$}
\author{Ke Wang$^{1}$}

\affiliation{$^{1}$State Key Laboratory of New Ceramic Materials, School of Materials Science and Engineering, Tsinghua University, Beijing, P. R. China.}
\affiliation{$^{2}$State Key Laboratory of Information Photonics and Optical Communications, School of Physical Science and Technology, Beijing University of Posts and
Telecommunications, Beijing, PR China}
\affiliation{$^{3}$Sichuan Zoige Alpine Wetland Ecosystem National Observation and Research Station, Sichuan Province Key Laboratory of Information Materials, Southwest Minzu University, Chengdu, P. R. China.}
\affiliation{$^{4}$Helmholtz-Zentrum Berlin für Materialien und Energie, Hahn-Meitner Platz 1, Berlin, Germany.}
\affiliation{$^{5}$Delft University of Technology, Mekelweg 15, Delft, The Netherlands.}
\affiliation{$^{6}$School of Physics, Southeast University, 211189 Nanjing, P. R. China.}
\affiliation{$^{7}$Wuzhen Laboratory, Jiaxing 314500, P. R. China.}
\affiliation{$^{8}$Institute of Materials, Swiss Federal Institute of Technology in Lausanne (EPFL), 1015 Lausanne, Switzerland.}
\affiliation{$^{9}$Graduate School, China Academy of Engineering Physics, Beijing, P. R. China.}

\date{\today}

\begin{abstract}
We report the discovery of a geometric pathway for tuning ferroelectric properties through thermally driven reconfiguration between coexisting polar states in Li-substituted NaNbO$_{3}$. Using first-principles density functional theory calculation and $^{7}$Li solid-state nuclear magnetic resonance spectroscopy measurement, we reveal that Li substitution creates two distinct polar configurations whose transformation under annealing enhances the Curie temperature and induces piezoelectric hardening. Our findings establish a geometrically-driven polar state reconfiguration mechanism, providing a general design principle for ferroics whereby macroscopic functional properties can be engineered via lattice geometry.
\end{abstract}

\maketitle

The geometry of the atomic lattice plays a decisive role in determining the physical properties of diverse material systems. In magnetism, geometric frustration prevents spins from simultaneously satisfying all preferred exchange interactions, giving rise to disordered or noncollinear ground states~\cite{RN301}. In superconductors, geometric effects manifest as lattice-controlled tuning of electronic bandwidth and pairing interactions~\cite{RN299,RN300}. In topological materials, lattice symmetry and structural geometry define the global topology of electronic wavefunctions~\cite{RN302}.

Conventional ferroelectricity arises from soft-mode instabilities induced by electronic hybridization between cations and anions~\cite{RN228}. However, ferroelectric instability can also be induced by geometric factors, \textit{e.g.}, rotation, distortions, or trimerization of coordination polyhedra~\cite{RN303, RN304, RN305}. Sodium niobate and lithium niobate exemplify these two contrasting mechanisms. In NaNbO$_{3}$, ferroelectricity arises from Nb-O covalent hybridization, characteristic of electronically driven ferroelectrics~\cite{RN217}. In contrast, LiNbO$_{3}$ exhibits polarization primarily due to geometric effects~\cite{RN291,RN308}, where large NbO$_{6}$ octahedral distortions induced by the small Li ion~\cite{RN309} trigger ferroelectricity independent of electronic hybridization. The interplay between these two mechanisms, \textit{i.e.}, electronic hybridization and geometric effect, defines a rich landscape of structural instabilities and functional tunability in perovskites.

Two decades ago, Kimura \textit{et al.} discovered the simultaneous enhancement of Curie temperature ($T_\mathrm{C}$) and piezoelectric hardening effect in lithium sodium niobate solid solution upon annealing, which are particularly valuable for high-temperature piezoelectric applications~\cite{RN214, RN215}. Specifically, annealing Li$_{0.06}$Na$_{0.94}$NbO$_{3}$ at sub-$T_\mathrm{C}$ of \SI{340}{\celsius} for 16~hours raised $T_\mathrm{C}$ from \SI{350}{\celsius} to \SI{425}{\celsius} and the mechanical quality factor from $\sim 500$ to $\sim 2500$~\cite{RN213}. Despite the significance of this discovery, the underlying mechanism remained unclear. Existing explanations include chemical element redistribution~\cite{RN275,RN289} and secondary-phase precipitation~\cite{RN276,RN277}, but none could account for the simultaneous enhancement of $T_\mathrm{C}$ and piezoelectric hardening under sub-$T_\mathrm{C}$ thermal treatment.

In this Letter, by combining first-principles density functional theory (DFT) calculations and \textsuperscript{7}Li solid-state nuclear magnetic resonance (NMR) spectroscopy, we uncover the atomistic origin of the annealing effect in ferroelectric Li-substituted NaNbO$_{3}$ and identify it as a manifestation of geometrically-driven reconfiguration between coexisting polar states. This transformation represents a reorganization mechanism of the polar configurations governed purely by geometric constraints, providing a new route to tune ferroelectric properties.

\begin{figure*}
    \includegraphics[width=14cm]{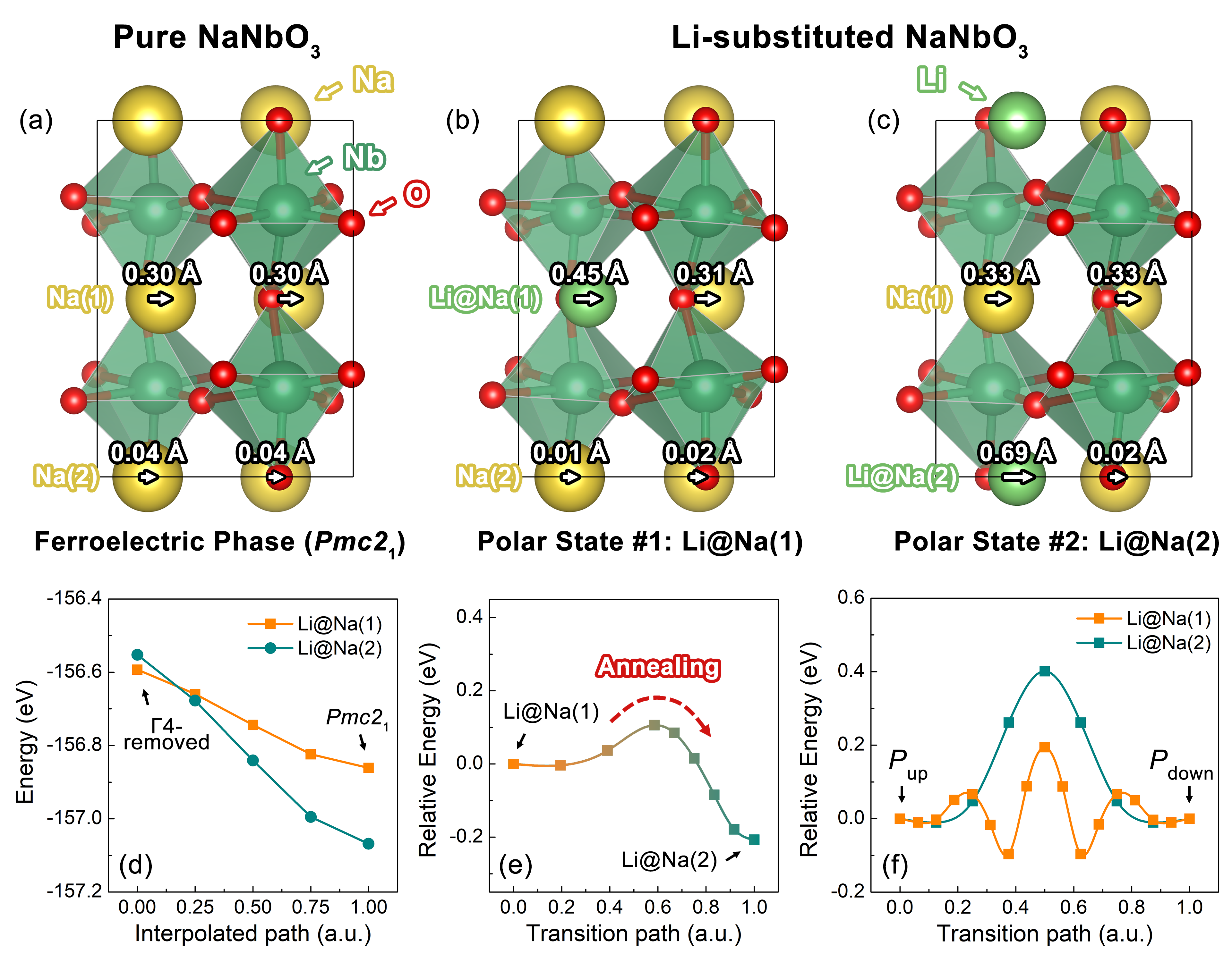}
    \caption{(a) DFT-simulated ferroelectric phase of pure NaNbO$_3$ with $Pmc2_1$ symmetry. Relaxed structures when Li cation occupies the (b) Na(1) or (c) Na(2) positions of NaNbO$_3$. (d) Interpolated structural variation of the Li-substituted NaNbO$_3$ between non-polar and polar states. (e) NEB-simulated transformation between two polar states. (f) NEB-simulated polarization switching in different polar states.}
\end{figure*}

NaNbO$_{3}$ is a complex ferroic system characterized by competing order parameters, such as polarization and octahedral tilting~\cite{RN279}. The energy proximity between the ferroelectric and antiferroelectric phases often results in their coexistence in synthesized materials~\cite{RN280, RN294}. Previous studies have demonstrated that minor Li substitution favors the formation of the ferroelectric phase in NaNbO$_{3}$, reasonably indexed to the $Pmc2_1$ symmetry~\cite{RN281}. As shown in Fig.~1(a), the ferroelectric $Pmc2_1$ phase of NaNbO$_{3}$ contains two distinct A-sites, \textit{i.e.}, Na(1) and Na(2). The off-centering displacement of Na(1) is 0.30~\AA{}, whereas Na(2) shows a much smaller displacement of 0.04~\AA{}. When Li substitutes at these sites, two unique local polar configurations emerge, denoted as Li@Na(1) and Li@Na(2) polar states. DFT structural relaxations [Figs.~1(b)-(c)] show that Li@Na(1) exhibits an off-centering displacement of 0.45~\AA{}, whereas Li@Na(2) displays an even larger displacement of 0.69~\AA{}, both exceeding the Na displacements 0.30~\AA{} and 0.04~\AA{} in pure NaNbO$_3$, respectively.

This asymmetry originates from the distinct AO$_{12}$ cuboctahedron environments around Na(1) and Na(2) with alternating obtuse and acute corners, which is a consequence of the specific tilting system ($a^-a^-c^+$) in $Pmc2_1$, as shown in Fig.~2. In the obtuse corner, the Na(1) cation is closer to the shared oxygen of NbO$_{6}$ octahedra (2.44~\AA{}), while in the acute corner, the distance increases to 2.87~\AA{} for Na(2) due to the steric hindrance from other neighboring oxygens. The smaller Li ion (0.76~\AA{} upon a coordination number of~6)~\cite{RN249} can move more freely toward the shared oxygen, shortening the distance to 2.06 and 2.09~\AA{} (for Li@Na(1) and Li@Na(2)) and thus enabling larger off-centering displacements. The octahedral tilting, and particularly the formation of the acute corner, introduces an additional geometric degree of freedom that allows A-site cations to undergo a larger off-centering displacement, especially the Li@Na(2) cation. Na(2) cannot take advantage of this geometric degree of freedom, owing to its larger ionic size (1.39~\AA{} upon a coordination number of~12)~\cite{RN249}, and instead experiences a constraining effect from it.

The tilting geometry creates a built-in asymmetry that allows two energetically distinct local minima, each corresponding to a unique polar state. DFT calculations reveal that the Li@Na(2) polar state ($-157.068$~eV/) is energetically more stable than the Li@Na(1) polar state ($-156.861$~eV/), as shown in Fig.~1(d). To understand the formation preference of these two polar states in Li-substituted NaNbO$_{3}$ upon cooling down from $T_\mathrm{C}$, it is necessary to investigate their non-polar state. To obtain this non-polar configuration, we remove the $\Gamma_4^-$ polar mode~\cite{RN306} from the structures of the relaxed $Pmc2_1$ Li@Na(1) and Li@Na(2) polar states using \textsc{ISODISTORT}~\cite{RN283}. Fig.~1(d) shows the energy difference between the Li@Na(1) and Li@Na(2) non-polar states to be very small ($\sim 0.04$~eV). This energetic proximity implies that both Na sites can be almost equally occupied by Li upon cooling, leading to the coexistence of polar states in the ferroelectric phase, which is an intrinsic feature of geometry.

\begin{figure}
    \includegraphics[width=7cm]{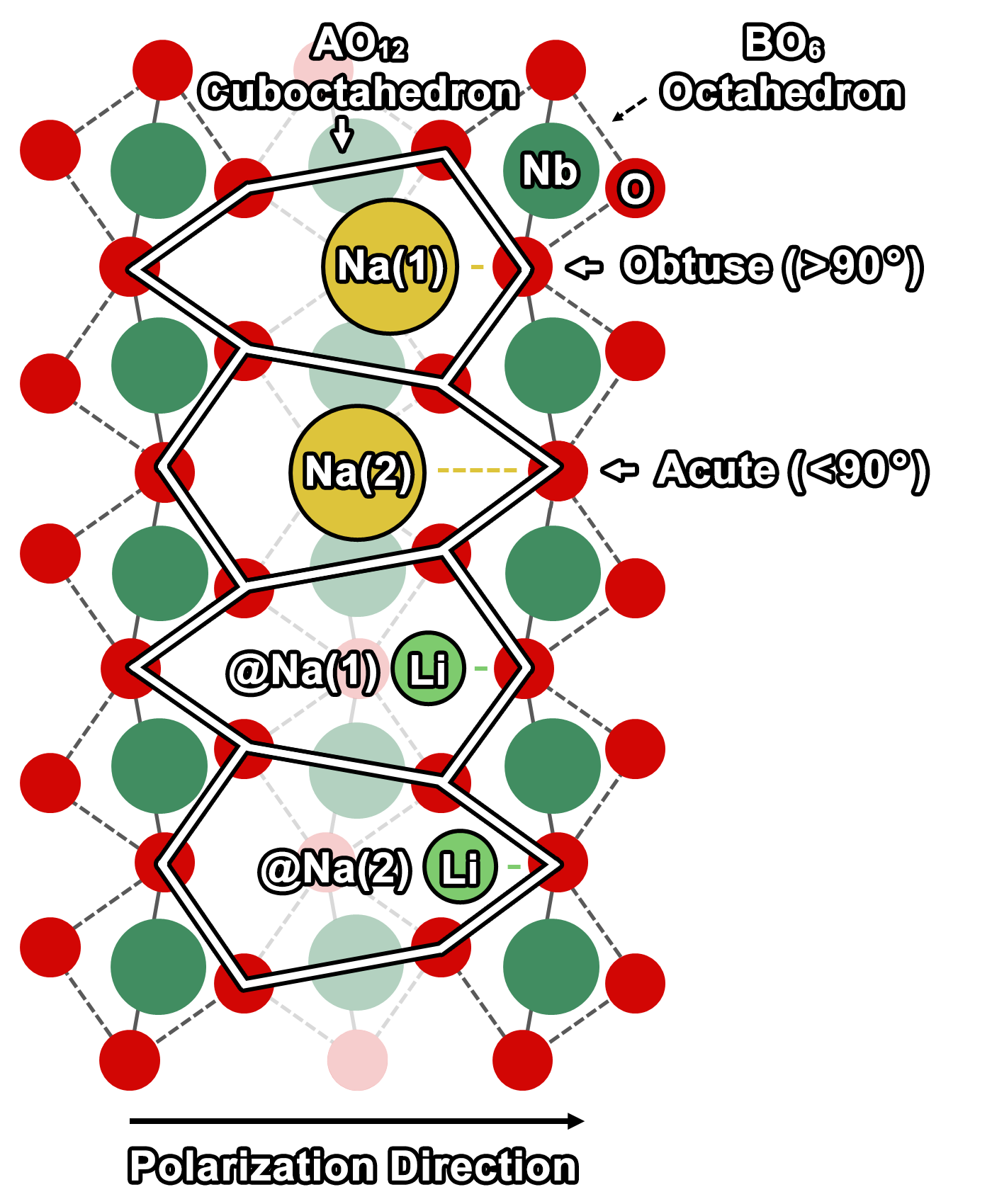}
    \centering
    \caption{ Illustration of the geometric constraint from octahedral tilting to the displacements of A-site cations. “Acute” and “obtuse” are defined according to the angle between two tilting BO$_6$ octahedra.}
\end{figure}

It is known that $T_\mathrm{C}$ is highly correlated with the polar-mode amplitude according to soft-mode theory.\cite{RN311} The geometry-allowed large displacement of Li ions enhances the polar mode, acting in concert with the conventional hybridization-driven ferroelectric instability. Based on the comparison of displacement amplitudes, it is reasonable to predict that the Li@Na(2) polar state should lead to a higher $T_\mathrm{C}$ than Li@Na(1). Our recent DFT investigation also revealed that Li incorporation at the A-site of potassium niobate enhances $T_\mathrm{C}$ through an enormous Li displacement up to 1\AA{} within the lattice, which amplifies the polar mode.\cite{RN278}

Nudged elastic band (NEB) calculations\cite{RN196} shown in Fig.~1(e) identify a small energy barrier separating the Li@Na(1) and Li@Na(2) polar states, corresponding to a displacive transformation pathway involving inverse octahedral tilting and cooperative atomic displacements (Fig.~3). We conjecture that this barrier is low enough to be overcome by mild thermal activation at sub-$T_\mathrm{C}$ of \SI{340}{\degreeCelsius}, coinciding with experimental annealing conditions.\cite{RN213} Annealing at sub-$T_\mathrm{C}$ thus drives a polar state reconfiguration from Li@Na(1) to Li@Na(2), shifting the population balance toward the state with higher $T_\mathrm{C}$. We further believe that the energy barrier between the Li@Na(1) and Li@Na(2) polar states is gradually minimized as the energy difference in interpolated configurations toward their nonpolar states decreases [Fig.~1(d)]. Both the lowered energy barrier and provided thermal energy are important ingredients of the annealing-driven $T_\mathrm{C}$ enhancement effect. 

\begin{figure}
    \includegraphics[width=9 cm]{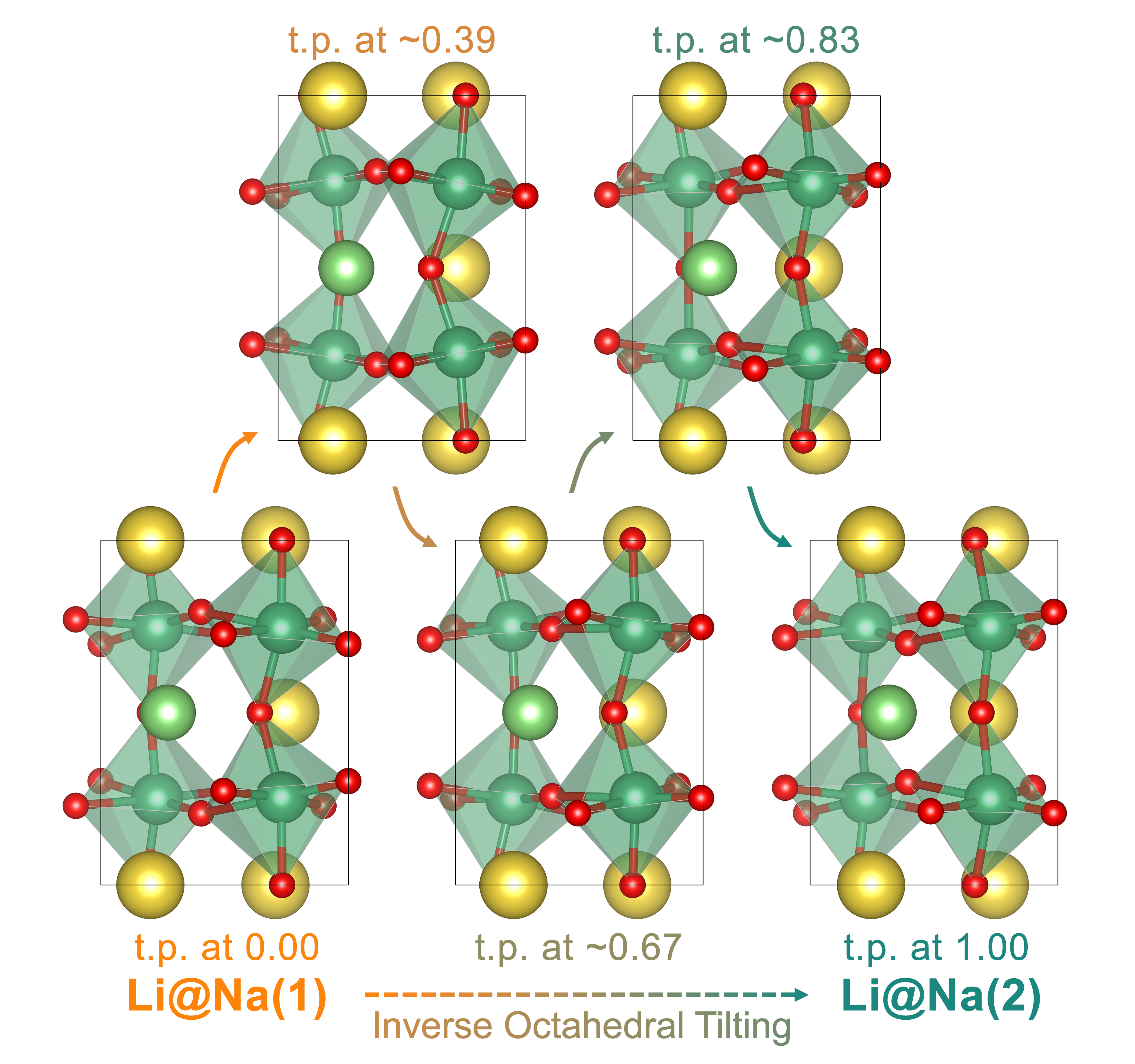}
    \centering
    \caption{Inverse octahedral tilting during the transformation from the Li@Na(1) polar state to the Li@Na(2) polar state. The “t.p.” is the abbreviation of transition path, obtained from the NEB simulation in Fig. 1(e).}
\end{figure}

\begin{figure*}
    \centering
    \includegraphics[width=12cm]{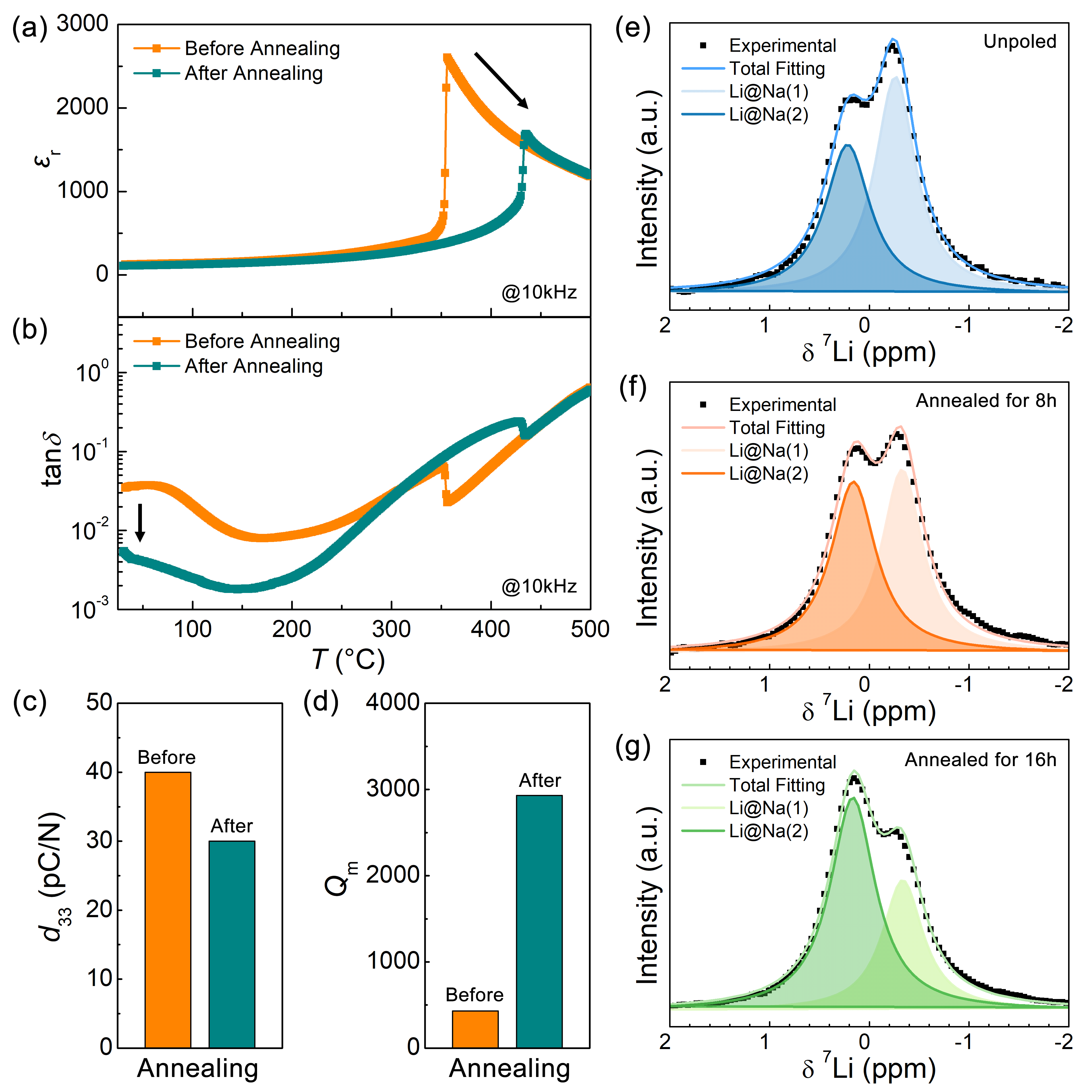}
    \caption{Temperature-dependent (a) dielectric permittivity $\varepsilon_r$, (b) loss $\tan\delta$, (c) piezoelectric constant $d_{33}$, and (d) mechanical quality factor $Q_m$ of Li-substituted NaNbO$_3$ ceramic samples before and after annealing. 
    $^{7}$Li NMR spectra of (e) unpoled and unannealed powder sample, (f) ceramic sample poled and annealed for 8 hours, and (g) ceramic sample poled and annealed for 16 hours.}
\end{figure*}

The inverse octahedral tilting during polar-state reconfiguration arises because the Na(1) and Na(2) sites differ by half a lattice periodicity. Inverse tilting redefines the topology of neighboring NbO$_6$ octahedra, effectively reconfiguring the Li coordination geometry and polar displacement. This geometrically-driven polarization reconfiguration is thus an intrinsic, lattice-controlled mechanism requiring only thermal energy sufficient to surmount a small internal barrier. It may be interesting to observe this reconfiguration using molecular dynamics simulations, especially with recent advances in machine-learning interatomic potentials that improve both accuracy and efficiency, e.g., in KNbO$_3$ and related solid solutions.\cite{RN286, RN287} We emphasize that the inverse octahedral tilting is a self-consistent explanation from DFT perspective consistent with experimental results, but there could be other possible mechanisms accounting for the phenomenon, which require further investigation. For example, the interdiffusion between Li and Na ions while maintaining the original octahedral tilting pattern could be another possible mechanism, though less likely due to the low sub-$T_\mathrm{C}$ annealing temperature.

The polarization switching behavior of both polar states, simulated via NEB [Fig.~1(f)], further supports this mechanism. The energy barrier for 180$^{\circ}$ polarization reversal in the Li@Na(2) polar state is approximately twice that of the Li@Na(1) polar state, indicating deeper potential wells and reduced domain-wall mobility. Consequently, as annealing drives the transformation toward the Li@Na(2) polar state, the material becomes intrinsically harder in terms of piezoelectricity, exhibiting higher coercive fields and mechanical quality factors but smaller piezoelectric coefficients, consistent with experimental observations [Figs.~4(a)–(d)]. Preparation details of ceramic samples can be found elsewhere.\cite{RN213}

The hardening thus originates not from conventional extrinsic defect pinning\cite{RN139} but from intrinsic lattice stiffening induced by geometric stabilization of a stronger polar state. This mechanism parallels the intrinsic hardening observed in tetragonal PbTiO$_3$-rich Pb(Zr,Ti)O$_3$ compositions,\cite{RN285} where deeper energy wells naturally yield larger coercive fields and lower losses. 

These predictions from simulations are supported by experimental evidence from NMR, as a powerful tool for identifying local ion environments. NMR observables such as chemical shifts and couplings depend directly on the immediate electronic and magnetic surroundings of specific nuclei, making it an element- and site-selective probe of short-range structure.\cite{RN293} Here, the coexistence and thermally driven reconfiguration of polar states are confirmed by $^7$Li NMR spectroscopy [Figs.~4(e)–(g)]. The spectrum in Fig.~4(e) exhibits two distinct resonances at $-0.3$~ppm and $+0.2$~ppm, corresponding respectively to Li@Na(1) and Li@Na(2) (see Appendix). Quantitative analysis of peak distribution (including centerband and spinning sidebands, see Appendix) shows a Li@Na(1)/Li@Na(2) peak-area ratio of 49:51, indicating that both polar states are equally populated in the virgin sample. As the sample is poled and annealed, the proportion of the Li@Na(2) polar state gradually increases with annealing duration, reaching 44:56 and 31:69 after 8~h and 16~h, respectively [Figs.~4(f)-(g)]. The excellent agreement between DFT predictions and NMR observations confirms their coexistence and transformation mechanism, validating the concept of geometry-governed polar-state reconfiguration as the microscopic origin of the annealing effect.

From the peak-area analysis, we note that the Li@Na(1) polar state does not completely transform into Li@Na(2) even after 16~h of annealing. Experimentally, further annealing yields no additional enhancement on $T_\mathrm{C}$ and hardening effect.\cite{RN213} The pronounced macroscopic changes despite incomplete transformation may originate from competition among different nuclei. During high-temperature sintering, Li is randomly distributed over both sites. Upon annealing, transformation from Li@Na(1) to Li@Na(2) may nucleate anywhere in the sample. Because the transformation involves inverse octahedral tilting, neighboring octahedra must rotate cooperatively, forming chain-like growth. Octahedral tilting, being an order parameter requiring an even number of lattice units, causes nuclei from random locations to eventually impinge if their tilting patterns are incompatible, producing interfacial defects that require higher energy to form. An equilibrium state emerges from competition between the energy-lowering reconfiguration and the energy-raising defect formation. Formation of finer domain configurations has been experimentally observed,\cite{RN213} likely representing an equilibrium structure associated with nucleation and growth within virgin domains.

The results presented here establish a unified geometrical mechanism for thermal tuning of ferroelectric properties. In Li-substituted NaNbO$_3$, the octahedral-tilting geometry creates multiple A-site environments supporting distinct polar configurations. Thermal energy near $T_\mathrm{C}$ enables reconfiguration favoring the state with a stronger polar mode. This geometrically-driven reconfiguration likely represents a general phenomenon applicable to other perovskites combining multiple cation environments and tilt instabilities. Chemically modified systems such as PbZrO$_3$-based solid solutions, which possess distinct A-site environments and strong tilting, are natural candidates for realizing similar geometry-enabled transformations. Research should not be limited to the $T_\mathrm{C}$ enhancement or hardening effect but extend toward engineering diverse functional properties. This geometric framework resolves the long-standing puzzle of annealing effects in alkali niobates and establishes a design principle applicable to a broad range of ferroelectrics. More generally, manipulating local octahedral-tilting geometry or cation-site occupancy provides a route to thermally reconfigurable ferroic states, which is a new degree of freedom for materials design.

In conclusion, this work uncovers a previously unexplored mechanism of ferroelectric engineering through thermally driven polar-state reconfiguration in Li-substituted NaNbO$_3$. The combination of insight from DFT simulations with experimental proof from NMR spectroscopy reveals that octahedral-tilting geometry and unique cation occupancy give rise to two coexisting polar states whose interconversion governs both $T_\mathrm{C}$ enhancement and intrinsic piezoelectric hardening. The findings reveal how local lattice geometry can act as a tunable order parameter in ferroic systems. More broadly, geometrically-driven polarization reconfiguration extends ferroelectric theory beyond the conventional paradigm, establishing a geometry-based framework for property design in functional materials.

\begin{acknowledgments}
K. W. acknowledges the support from National Natural Science Foundation of China (No. 52325204, 52421001). H.-C. T. acknowledges the support from National Natural Science Foundation of China (No. W2433118). M.-H. Z acknowledges the support from the National Natural Science foundation of China (No. 12574103), the Zhejiang Provincial Leading Innovation and 
Entrepreneurship Team Project (2025R01021), and the National Key R\&D Program of China (No. 2024YFF1400702). B. W. acknowledges the support from National Nature Science Foundation of China (No. 12474091), the Foundation of Sichuan Province Science and Technology Support Program (No. 2025NSFJQ0026). P. G. acknowledges financial support by the Dutch Research Council (NWO) for the ECCM Tenure Track funding (No. ECCM.006) and the German Federal Ministry of Education and Research (No. 03SF0565A).  The authors would like to acknowledge Prof. Philippe Ghosez from University of Liège and Prof. Zhi Tan from Sichuan University for insightful discussions. 
\end{acknowledgments}

\nocite{*}

\bibliography{main}

\onecolumngrid
\clearpage

\begin{center}
    \textbf{\large End Matter}
\end{center}
\vspace{10pt}
\twocolumngrid

\setcounter{section}{0}
\renewcommand{\thesection}{Appendix~\Alph{section}}
\appendix

\section{DFT Calculation}
Density functional theory (DFT) calculations were performed using the Vienna Ab initio Simulation Package (VASP) with the projector-augmented wave (PAW) method. The exchange-correlation energy was treated within the Perdew-Burke-Ernzerhof (PBE) generalized gradient approximation (GGA). A plane-wave cutoff energy of 500 eV was employed, and the Brillouin zone was sampled using \( \Gamma \)-centered k-point meshes with spacing ~0.12 Å\textsuperscript{-1}. Convergence criteria for the electronic self-consistent field loop were set to 10\textsuperscript{-6} eV, while atomic positions were relaxed until the residual Hellmann–Feynman forces fell below 0.005 eV/Å.

\section{Li Concentration}
The concentration of Li in NaNbO$_3$ is investigated, as shown in Fig.~5. Different Li concentrations are considered, including 25\%, 12.5\%, and 6.25\%, using 20, 40, and 80 atoms for constructing the supercells, respectively. An energy difference exists between Li@Na(1) and Li@Na(2) polar states, regardless of the concentration. The energy difference as a function of concentration remains roughly constant, suggesting that the impact of Li substitution is predominantly short-range. Namely, having Li closer to each other does not lead to any other significant physical mechanism that potentially changes the energy. From this perspective, simulations using a 20-atom supercell (25\% of Li) are methodologically reasonable, as Li-Li interactions remain weak enough to be ignored.

\begin{figure}
    \includegraphics[width=9 cm]{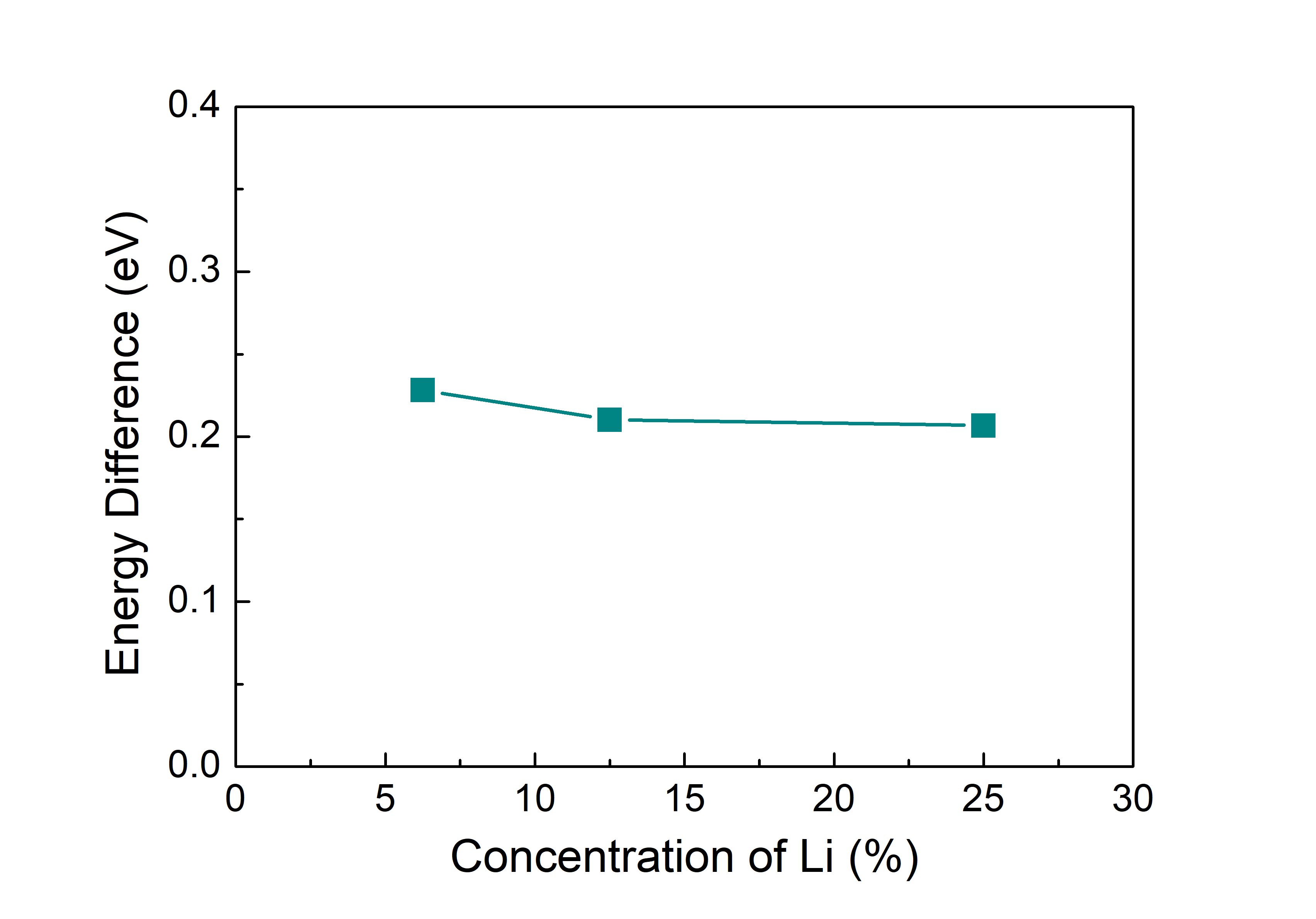}
    \centering
    \caption{Energy difference between Li@Na(1) and Li@Na(2) polar states as a function of Li concentration.}
\end{figure}

\section{\textsuperscript{7}Li Chemical Shift Simulation}
The calculation of \textsuperscript{7}Li chemical shifts from DFT at both Li@Na(1) and Li@Na(2) sites is performed by using CASTEP, and found to be -90.35 and -89.38 ppm, respectively. According to the empirical 1-to-1 correspondence between simulated chemical shift and experimental chemical shift,\cite{D4TA04628E} the simulated values can be roughly equal to -0.42 and 0.30 ppm in the experimental scale. The simulated chemical shift of Li@Na(1) is more negative than that of Li@Na(2), qualitatively consistent with the experimental trend of a more shielded environment for Li@Na(1), but the calculated values deviate from experiment. This might originate from the fact that the functional used in DFT is only an approximation.

\section{NMR Measurement}
$^7$Li NMR spectra were recorded with a Bruker Ascend 400 magnet ($B_0 = 9.4~\mathrm{T}$) equipped with a NEO console operating at a resonance frequency of 155.52~MHz. The chemical shift was referenced with respect to LiCl in the solid state, which exhibits a resonance at $-1.1~\mathrm{ppm}$. A solid $\pi/6$ pulse length of 1~$\mu$s was used for data acquisition. The sample was spun at a frequency of 3~kHz in a 4~mm rotor using a Bruker two-channel magic-angle spinning (MAS) probe. A total of 32 scans were recorded with a recycle delay of 60~s using a single-pulse sequence, thereby ensuring full thermal equilibrium of the nuclear spin population before each scan.

\begin{figure*}
    \includegraphics[width=14 cm]{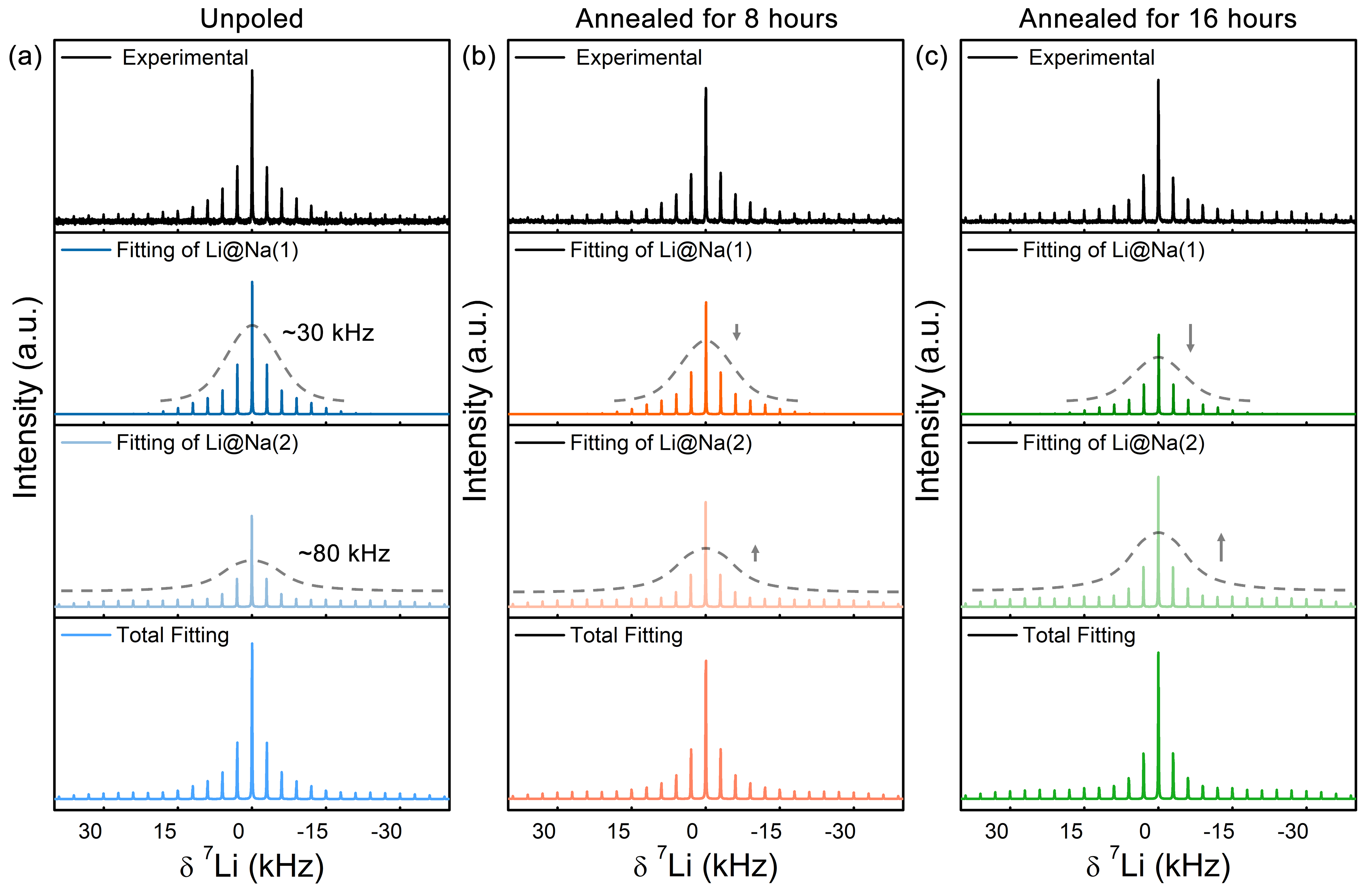}
    \centering
    \caption{$^7$Li MAS NMR full spectra of (a) unpoled and unannealed powder sample, (b) ceramic sample poled and annealed for 8 hours, and (c) ceramic sample poled and annealed for 16 hours. Grey dashed lines are drawn above the fitting curves of Li@Na(1) and Li@Na(2) to guide the eye. }
\end{figure*}

All peaks of the experimental spectrum (centreband and spinning sidebands) at 3~kHz MAS were fitted to two components, corresponding to either the Li@Na(1) or Li@Na(2) polar state, respectively. For consistency, the distance between spinning sidebands of a given component was ensured to match the MAS frequency. Next, the fitted spinning sideband pattern for each component was extracted and simulated separately using the \textsc{DMFIT} program~\cite{RN288} to determine the NMR parameters for each individual site. 

Using the individual sets of NMR parameters, including line shape (Gaussian–Lorentzian) and line width, $C_Q$, $\eta$, isotropic chemical shift, dipolar component, etc., the complete spectrum was fitted to a weighted distribution based on the two polar states. The goodness of fit is summarized in Table~1. Subsequently, the area of each pattern was integrated for further analysis. 

\begin{table}[!h]
\caption{Summary of the NMR spectra fitting statistics.}
\begin{ruledtabular}
\begin{tabular}{lccc}
\textbf{Statistics} & \textbf{Unpoled} & \textbf{Annealed} & \textbf{Annealed} \\ & & \textbf{for 8 h} & \textbf{for 16 h}\\
\hline
$R^2$ & 0.97 & 0.98 & 0.98 \\
Reduced $\chi^2$ & 1.26 & 1.45 & 1.47 \\
\end{tabular}
\end{ruledtabular}
\end{table}

\begin{table}[!h]
\caption{Relative area proportion of the Li@Na(1) and Li@Na(2) fitting curves averaged from 20 Monte Carlo–based fitting runs of the full spectrum (including centerband and spinning sidebands). The parentheses show the standard deviation.}
\begin{ruledtabular}
\begin{tabular}{lccc}
\textbf{Model} & \textbf{Unpoled} & \textbf{Annealed} & \textbf{Annealed} \\ & & \textbf{for 8 h} & \textbf{for 16 h}\\
\hline
Li@Na(1) & 49.13\% (0.13\%) & 43.96\% (0.12\%) & 31.00\% (0.09\%) \\
Li@Na(2) & 50.87\% (0.13\%) & 56.04\% (0.12\%) & 69.00\% (0.09\%) \\
\end{tabular}
\end{ruledtabular}
\end{table}

A side note should be mentioned on the only partially averaged dipolar coupling for $^7$Li (usually of a magnitude of 9-10~kHz) while spinning at 3~kHz, as this interaction and its effect on the spinning sidebands from the central transition were reproduced with a non-axially symmetric chemical shift anisotropy (CSA) tensor of about 20~ppm span.

Evaluating the MAS spectra at this low rotation frequency provides insight into the quadrupolar coupling constant ($C_Q$) for each site (see Fig.~6). The spinning sidebands envelope for each Li site was extracted and fitted individually, ultimately using the extracted NMR values to fit the combined spectrum. In detail, while the peak at $+0.2~\mathrm{ppm}$ exhibits a $C_Q \approx 80~\mathrm{kHz}$, the signal at $-0.3~\mathrm{ppm}$ displays a $C_Q \approx 30~\mathrm{kHz}$. 

As the quadrupolar coupling constant $C_Q$ provides a measure of the local distortion of a particular site, it is reasonable to assert that the site at $+0.2~\mathrm{ppm}$ with $C_Q \approx 80~\mathrm{kHz}$ is located in a more distorted environment than the site at $-0.3~\mathrm{ppm}$ with $C_Q \approx 30~\mathrm{kHz}$. Regarding the assignment of both Li resonances, the larger $C_Q \approx 80~\mathrm{kHz}$ must correspond to the Li site with the larger cationic displacement (\textit{i.e.}, Li@Na(2)). 

In detail, the $C_Q$ value is a parameter mainly defined by the spatial distribution of charges in the immediate surroundings of the probed nucleus and is directly proportional to the distortion of the local environment and the electric field gradient (EFG). Furthermore, in contrast to the Na sites, which exhibit very distinct asymmetry parameters $\eta$ for the EFG tensor, the spinning sidebands envelope for both Li sites could be best fitted with an asymmetry parameter $\eta = 1$. This value reflects a non-axially symmetric environment for both Li sites, which is reasonable given the smaller cation radius of Li compared to Na, and the consequent ion displacement from the center of the AO$_{12}$ polyhedron that would occur for either Li site upon substitution.

\end{document}